\documentclass[%
 aip,
 jmp,%
 amsmath,amssymb,
 reprint,%
]{revtex4-1}
\usepackage{amsfonts}
\usepackage{dcolumn}% Align table columns on decimal point
\usepackage{amsmath}
\usepackage{amssymb}
\usepackage{graphicx}
\usepackage{float}

\begin{document}
\title{Spin transport in an insulating ferrimagnetic-antiferromagnetic-ferrimagnetic trilayer as a function of temperature}

\author{Yizhang Chen}
\affiliation{Center for Quantum Phenomena, Department of Physics, New York University, New York, New York 10003, USA}

\author{Egecan Cogulu}
\affiliation{Center for Quantum Phenomena, Department of Physics, New York University, New York, New York 10003, USA}

\author{Debangsu Roy}
\affiliation{Center for Quantum Phenomena, Department of Physics, New York University, New York, New York 10003, USA}

\author{Jinjun Ding}
\affiliation{Department of Physics, Colorado State University, Fort Collins, Colorado 80523, USA}

\author{Jamileh Beik Mohammadi}
\affiliation{Center for Quantum Phenomena, Department of Physics, New York University, New York, New York 10003, USA}

\author{Paul G. Kotula}
\affiliation{Sandia National Laboratories, Albuquerque, New Mexico 87185, USA}

\author{Nancy A. Missert}
\affiliation{Sandia National Laboratories, Albuquerque, New Mexico 87185, USA}

\author{Mingzhong Wu}
\affiliation{Department of Physics, Colorado State University, Fort Collins, Colorado 80523, USA} 

\author{Andrew D. Kent}
\email{andy.kent@nyu.edu}
\affiliation{Center for Quantum Phenomena, Department of Physics, New York University, New York, New York 10003, USA}
\date{\today}


\begin{abstract}
We present a study of the transport properties of thermally generated spin currents in an insulating ferrimagnetic-antiferromagnetic-ferrimagnetic trilayer over a wide range of temperature. Spin currents generated by the spin Seebeck effect (SSE) in a yttrium iron garnet (YIG) YIG/NiO/YIG trilayer on a gadolinium gallium garnet (GGG) substrate were detected using the inverse spin Hall effect in Pt. By studying samples with different NiO thicknesses, the NiO spin diffusion length was determined to be 4.2 nm at room temperature. Interestingly, below 30 K, the inverse spin Hall signals are associated with the GGG substrate. The field dependence of the signal follows a Brillouin function for a S=7/2 spin ($\mathrm{Gd^{3+}}$) at low temperature. Sharp changes in the SSE signal at low fields are due to switching of the YIG magnetization. A broad peak in the SSE response was observed around 100 K, which we associate with an increase in the spin-diffusion length in YIG. These observations are important in understanding the generation and transport properties of spin currents through magnetic insulators and the role of a paramagnetic substrate in spin current generation.
\end{abstract}
\maketitle

% Introduction
\section{Introduction}

A spin current, or a flow of spin angular momentum, can be carried by conduction electrons \cite{Fert_PRL1988, Grunberg_PRB1989} or spin waves \cite{Berger_PRB1996, Slonczewski_JMMM1999}. In a material with large spin-orbit coupling, like Pt, a spin current can be converted into a measurable voltage by the inverse spin Hall effect (ISHE) \cite{Kimura_PRL2007}. Spin currents can be generated by the spin Hall effect (SHE) \cite{Dyakonov_PLA1971, Hirsch_PRL1999, SZhang_PRL2000}, spin pumping \cite{spinPumping_PRL2002}, or the spin Seebeck effect \cite{Uchida_Nature2008, Jaworski_Nature2010, Uchida_APL2010, SSE_AFM}. The spin Seebeck effect refers to the generation of spin currents when a temperature gradient is applied to a magnetic material and has potential applications in converting waste heat into electricity.

% FIGURE 1
\begin{figure}[t]
\includegraphics[width=1\linewidth]{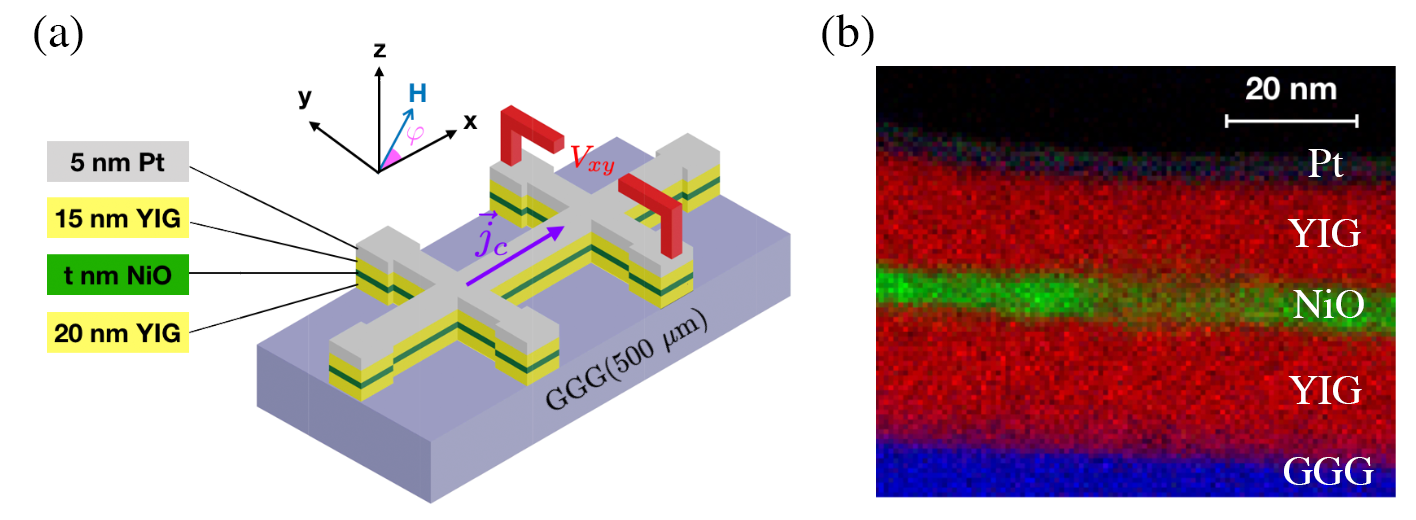}
\caption{A schematic of the sample and cross-sectional characterization of the sample by scanning transmission electron microscopy energy-dispersive X-ray spectroscopy (STEM-EDS). (a) Sample geometry showing the layers, the electrical contacts and the applied magnetic field. $\vec{j}_c$ is the density of the charge current applied in the $x$-direction.  $V_{xy}$ is the voltage measured in the transverse direction, and $\varphi$ is the angle between the applied magnetic field and the current. GGG, YIG, NiO, and Pt are represented as purple, yellow, green, and grey, respectively. (b) Sample cross section characterized by STEM-EDS. GGG, YIG, NiO, and Pt layers are colored in blue, red, green, and dark gray, respectively.}
\label{fig:fig1}
\end{figure}

A conventional spin valve consists of two ferromagnetic metals separated by a non-magnetic metal \cite{Parkin_JAP1991, Jedema_nature2001}. Recently, a new spin valve structure based on an antiferromagnetic insulator (AFI) sandwiched between two ferromagnetic insulators (FI) was proposed \cite{Shufeng_APL2018}. An AFI can conduct both up and down spins due to the degeneracy of its magnon spectrum at zero field. The predicted valve effect associated with thermally induced spin currents has been observed by controlling the relative orientations of $\mathrm{Y_3Fe_5O_{12}}$ (YIG) magnetization in a YIG/NiO/YIG structure \cite{Han_PRB2018}. YIG is a ferrimagnetic insulator with low magnetic dissipation, highly efficient spin current generation \cite{Heinrich_PRL2011}, and long-distance magnon transport \cite{Wesenberg_np2017}. Nickel Oxide (NiO) is an antiferromagnetic insulator used to decouple the two ferrimagnetic layers while conducting thermally generated spin currents.  

% FIGURE 2
\begin{figure*}[t]
\includegraphics[width=1\textwidth]{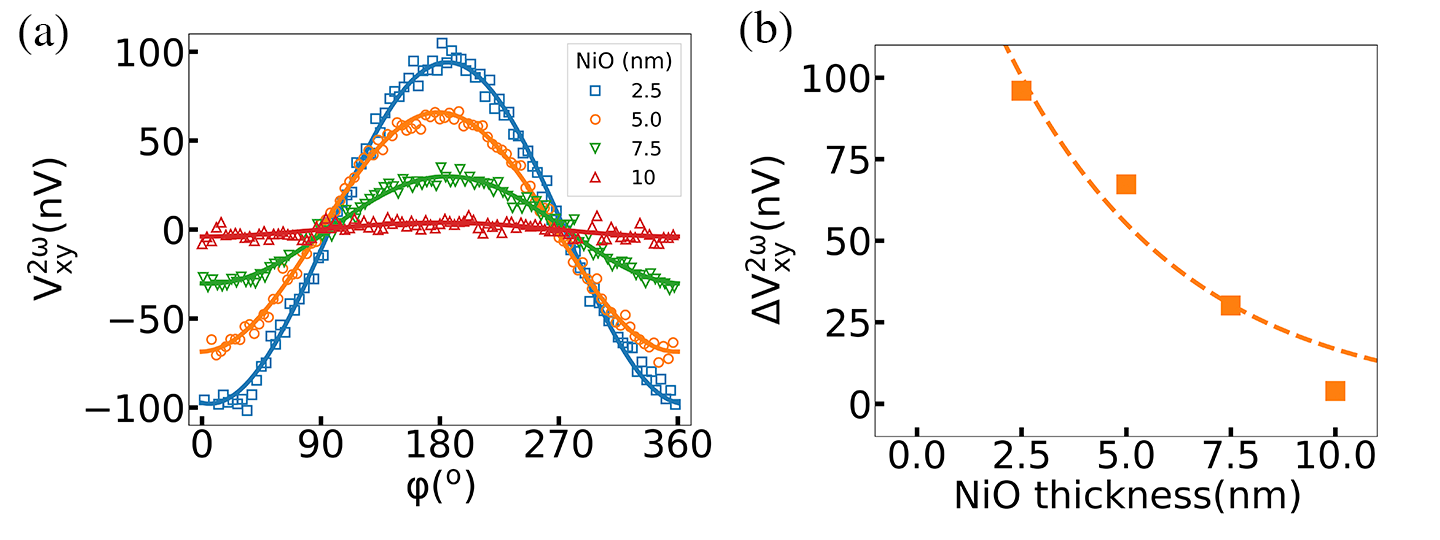}
\caption{Angular dependence and NiO thickness dependence of $V_{xy}^{2\omega}$ measured with an in-plane magnetic field of $0.4\ \mathrm{T}$ at room temperature. (a) Angular dependence of $V_{xy}^{2\omega}$ with an AC density of $j_{AC} = 1.5 \times 10^{10} \mathrm{A/m^2}$. $\Delta V_{xy}^{2\omega}$ is extracted by fitting the curve with a cosine function. (b) $\Delta V_{xy}^{2\omega}$ as a function of the NiO thickness. The curve is fitted with $V = V_0 e^{-t/\lambda_{\mathrm{NiO}}}$, where $V_0 = 182\ \pm 44\ \mathrm{nV}$ and the spin diffusion length of NiO is $\lambda_{\mathrm{NiO}} \approx 4.2 \pm 1.1\ \mathrm{nm}$.}
\label{fig:fig2}
\end{figure*}

To understand the generation, transmission, and detection of spin currents through a multilayer consisting of different magnetic insulators, transport measurements were performed in samples consisting of GGG(500 $\mathrm{\mu m}$)/YIG(20 nm)/NiO(t nm)/YIG(15 nm)/Pt(5 nm). Here GGG ($\mathrm{Gd_3Ga_5O_{12}}$) is the standard substrate used to grow epitaxial YIG. Above the spin-glass transition temperature ($\sim-0.18$ K), GGG is paramagnetic with no long-range magnetic order \cite{Schiffer_PRL1995, Koichi_arxiv2018}. In addition, GGG has been shown to have a SSE, with a magnitude comparable to the SSE that of YIG at low temperatures \cite{PM_SSE}.

In this article, room-temperature measurements were first performed to characterize the spin diffusion length of NiO. Then experiments were conducted over a broad range of temperature from 5 to 300 K. These revealed a strong enhancement of the SSE below 30 K that originates from the GGG substrate. Further, field-dependent experiments show behavior associated with switching of YIG magnetization and paramagnetism of GGG. Furthermore, a broad peak in the SSE response around 100~K was observed, which may originate from the temperature dependence of the spin diffusion length in YIG.

% Sample fabrication and measurement techniques
\section{Sample fabrication and measurement techniques}
The sample was fabricated in the following way. First, a 20 nm YIG layer was grown epitaxially on a (111)-oriented GGG substrate (500 $\mu \mathrm{\textbf{m}}$) at room temperature and annealed in $\mathrm{O_2}$ at high-temperature \cite {Chang_IEEE2014}. An Ar plasma was used to clean the surface of the samples before depositing NiO via radio frequency (RF) sputtering in another chamber. Afterward, a 15 nm YIG layer was grown on top with the same growth conditions of the first layer. Then the sample was capped with a 5~nm Pt layer. For transport and SSE measurements, the Pt was patterned into Hall bar structures using electron beam lithography and Ar plasma etching. The Hall bar has a width of 4 $\mu$m and the length between the two longitudinal contacts is 130 $\mu \mathrm{m}$. An alternating current (AC) with a frequency of 953 Hz was used. As the temperature gradient induced by the AC oscillates at twice the frequency, the second harmonic Hall voltage $V_{xy}^{2\omega}$ measured by a lock-in amplifier is proportional to the amplitude of the SSE-produced spin current \cite{Yizhang_APL2018, YChen_PhDThesis}. Room-temperature measurements were performed with a 0.4 T magnetic field applied in-plane. Temperature-dependent measurements are carried out in the Quantum Design PPMS system also with an in-plane applied magnetic field.

% Experimental Results
\section{Experimental Results}

% FIGURE 3
\begin{figure*}[t]
\includegraphics[width=1\textwidth]{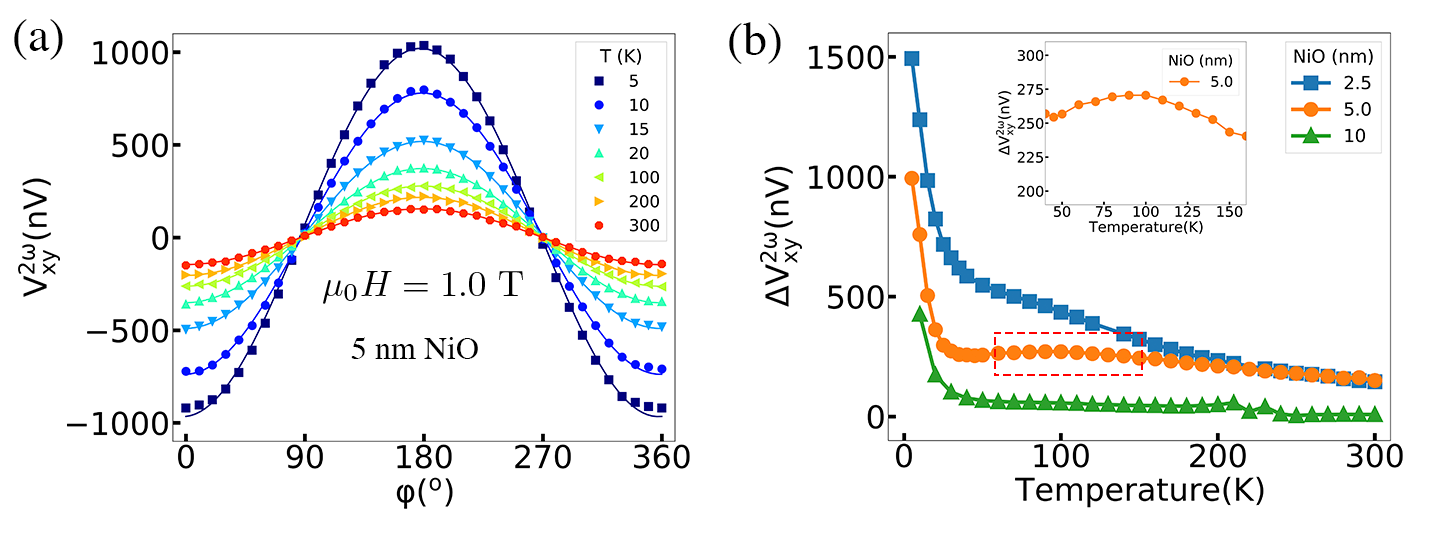}
\caption{Second harmonic response $V_{xy}^{2\omega}$ measured at several temperatures with an applied magnetic field $\mu _0 H = 1.0\ \mathrm{T}$ for NiO thicknesses of 2.5, 5, and 10 nm. (a) Angular dependence of $V_{xy}^{2\omega}$ measured from 5 to 300 K. Note that the angle $\varphi$ is defined in Fig. \ref{fig:fig1}(a). (b) $\Delta V_{xy}^{2\omega}$ measured as a function of the temperature. Inset: a broad peak is observed around 100 K for the sample with a 5 nm NiO thickness.}
\label{fig:fig3}
\end{figure*}

Figure \ref{fig:fig1}(a) is a  schematic of the GGG/YIG/NiO/YIG/Pt sample. A magnetic field is applied in-plane at an angle $\varphi$ with respect to the current. The cross section of the sample is characterized by scanning transmission electron microscopy with energy-dispersive X-ray spectroscopy, shown in Fig. \ref{fig:fig1}(b). Both the top and bottom YIG layers are crystalline, with thickness of 15 nm and 20 nm. NiO is polycrystalline, with a thickness of 5 nm for this sample (see Fig. S1 in the supplemental materials).

% NiO spin diffusion length 
%\textbf{2. Spin currents induced by spin Seebeck effect detected as the second harmonic responses for different NiO thickness.}

First, $V_{xy}^{2\omega}$ was measured as a function of $\varphi$ for samples with NiO thicknesses of 2.5, 5, 7.5, and 10 nm. $V_{xy}^{2\omega}$ reaches a maximum at $\varphi = 180^o$ and minimum at $\varphi = 0^o$. This is consistent with the ISHE symmetry of the Hall voltage $V_{\mathrm{ISHE}} \propto \vec{j}_s \times \hat{\sigma} \propto \nabla T \times \hat{m} \propto \cos(\varphi)$, where $\vec{j}_s$ is the spin current, $\hat{\sigma}$ is the spin polarization direction, $\nabla T$ is the temperature gradient, and $\hat{m}$ is a unit vector in the direction of magnetization. The angular dependence of the $V_{xy}^{2\omega}$ was fitted with a cosine function and the amplitude $\Delta V_{xy}^{2\omega}$ is plotted as a function of the NiO thickness (Fig. \ref{fig:fig2}). $\Delta V_{xy}^{2\omega}$ decays rapidly as the NiO thickness increases and is fitted to an exponentially decaying function $V = V_0 e^{-t/\lambda_{\mathrm{NiO}}}$. The characteristic spin diffusion length of NiO is $\lambda_{\mathrm{NiO}} \approx 4.2\pm 1.1\ \mathrm{nm}$, close to what has been found in previous work on YIG/NiO/Pt structures \cite{Yuming_AIP2017}.

% Temperature dependence
%\textbf{3. Temperature dependence of the second harmonic signals.} 

To further understand the generation and transport of thermally generated spin currents through the heterostructure, the angular dependence of $V_{xy}^{2\omega}$ was measured from 5 to 300 K with an applied magnetic field of 1.0 T (Fig. \ref{fig:fig3}(a)). The amplitude $\Delta V_{xy}^{2\omega}$ is extracted by the same method discussed above and is plotted as a function of the temperature (Fig.~\ref{fig:fig3}(b)). For the 5 nm thick NiO sample, as temperature decreases from 300 to 100~K, $\Delta V_{xy}^{2\omega}$ increases steadily from 150 to 271 nV. From 100 to 50 K, $\Delta V_{xy}^{2\omega}$ slightly decreases to 257~nV, forming a broad peak around 100~K, shown in the inset of Fig.~\ref{fig:fig3}(b). However, as temperature decreases below 30~K, $\Delta V_{xy}^{2\omega}$ increases dramatically from 297 to 994~nV. The enhancement below 30~K was observed for all samples.

%\section{4. Field dependence of SSE.}
As has been previously noted, the SSE depends on the magnon population, the spin diffusion length, and the interfacial spin-mixing conductance in the heterostructure. In order to understand the correlation between the SSE signal and the magnetization of the samples, field-dependent measurements of $V_{xy}^{2\omega}$ were performed in the sample with 2.5 nm thick NiO. Fig. \ref{fig:fig4}(a) shows $V_{xy}^{2\omega}$ as a function of the applied magnetic field between -5.0 and 5.0~T with temperature ranging from 5 to 50~K. At 50 K, as the applied field goes from -5.0 to -1.0 T, $V_{xy}^{2\omega} \approx 530\ \mathrm{nV}$, almost independent of the applied field. As the applied field increases from -1.0~T to 20 mT, $V_{xy}^{2\omega}$ decreases slowly to 108 nV. Then $V_{xy}^{2\omega}$ drops sharply to -156 nV as the applied field increases from 20 to 100 mT. As the applied field increases to 1.0~T, $V_{xy}^{2\omega}$ decreases slowly to -540 nV and again is nearly constant thereafter. The sharp switching steps observed in the $V_{xy}^{2\omega} - H$ curves around $\pm 50\ \mathrm{mT}$ occur at the coercive field of the YIG, which is smaller than 50 mT at room temperature (see Fig. S2 in the supplemental materials). Only one magnetization reversal can be identified between -200 and 200 mT in the $V_{xy}^{2\omega} - H$ curves. As temperature decreases from 50 to 5~K, $V_{xy}^{2\omega}$ increases from 535 to 1970 nV, while the low-field step does not change significantly. A clear correlation between the $V_{xy}^{2\omega}$ and the magnetization of a paramagnet can be seen by comparing the $V_{xy}^{2\omega}-H$ curves with the Brillouin function of a S = +7/2 spin ($\mathrm{Gd^{3+}}$), shown in Fig. \ref{fig:fig4}(b).

% Only one sharp reversal was observed, as the top and bottom YIG layers possess non-observable difference in coercivity (see supplement). The reversal of of $V_{xy}^{2\omega}$ at low field corresponds to the switching of top and bottom YIG magnetization. At 300 K, $V_{xy}^{2\omega}$ experiences little changes after the low-field switching. However, as temperature decreases, $V_{xy}^{2\omega}$ shows additional contribution in the high-field region. At 5 K, the high field contribution becomes dominant to $V_{xy}^{2\omega}$, which accounts for over 90\%. FIG. \ref{fig:fig4} (b) is the Brillouin function of a S = +7/2 spin ($\math{Gd^{3+}$) from 5 to 50~K. 

\section{Discussion}
\begin{figure*}[t]
\includegraphics[width=1\textwidth]{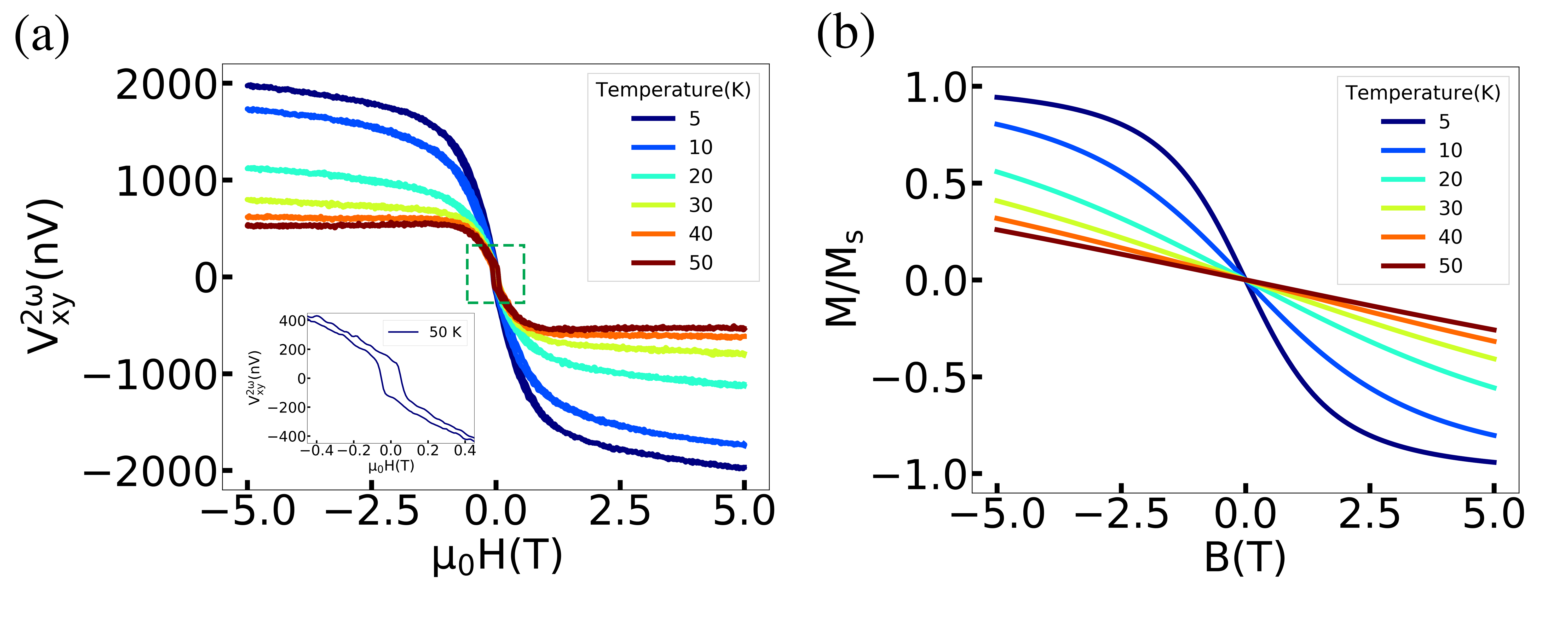}
\caption{(a) Field dependence of the $V_{xy}^{2\omega}$ measured between 5 and 50 K, with $\varphi = 0 ^o$. The sample is GGG (500 $\mu$m)/YIG(40 nm)/NiO(2.5 nm)/YIG(20 nm)/Pt (5 nm). The magnetic field is swept between -5 and +5 T. The offset of $V_{xy}^{2\omega}$ has been removed. (b) Brillouin function of an S=+7/2 spin.}
\label{fig:fig4}
\end{figure*}

The SSE voltages decay rapidly as NiO thickness increases, as presented in Fig. \ref{fig:fig2}. This indicates that spin currents were generated not only from the top YIG layer but also from the bottom YIG or GGG layer. The NiO spin diffusion length is close to what has been found before at room temperature in YIG/NiO/Pt structures \cite{Yuming_AIP2017}.

A dramatic enhancement of the SSE voltages has been observed below 30 K, which is likely associated with GGG. The same enhancement has been observed in GGG(500\ $\mu \mathrm{m}$)/YIG(20 nm)/Pt(5 nm), shown in Fig.~S3 in the supplemental materials. A previous study has shown that the spin current $j_{s}$ generated by paramagnetic SSE in GGG/Pt bilayer has a $T^{-1}$ temperature dependence, associated with GGG susceptibility, which follows the Curie-Weiss law $\chi = C/(T - \Theta_{CW}$), where $C$ is the Curie constant and $\Theta_{CW}$ Curie-Weiss temperature\cite{PM_SSE}. At low temperatures, the GGG thermal conductivity $k_{\mathrm{GGG}}$ has a $T^{3}$ temperature dependence. Therefore, the temperature gradient generated by a constant power is $\nabla T \propto 1/k_{\mathrm{GGG}} \propto 1/T^{-3}$. The resulting SSE voltage goes as $V_{SSE} \propto j_s \cdot \nabla T \propto T^{-4}$. In addition, a broad peak of the SSE signal observed around 100~K in the YIG/Pt structure suggests that the spin diffusion length in YIG has a strong temperature dependence \cite{Guo_PRX2016}. As spin currents generated and transmitted through YIG layers, the temperature-dependent spin diffusion length in YIG would have a significant effect on the ISHE voltage generated in Pt. So the broad peak observed around 100~K in Fig. \ref{fig:fig3} may be associated with the temperature dependence of spin diffusion length in YIG. However, further experiments are needed to understand how spin currents are transmitted through bulk GGG, NiO, GGG/YIG, YIG/NiO, and NiO/YIG interfaces at different temperatures.

%In addition, the spin diffusion length of YIG depends on temperature, shown in previous experiments \cite{Guo_PRX2016}. The pronounced peak observed around 100 K in Fig. \ref{fig:fig3} may be associated with the temperature dependence of spin diffusion length in YIG. 

Comparing the field dependence of SSE voltages and the Brillouin function from 5 to 50 K, it is clear that there is a contribution to SSE from GGG at low temperatures. At 5 K, the SSE voltage follows the Brillouin function as the magnetic field swept from -5.0 to 5.0 T. As temperature increases, the SSE voltages start deviating from the Brillouin function (Fig. \ref{fig:fig4} and Fig. S4). The underlying physics is not yet fully understood, since the role played by GGG, YIG, NiO and their corresponding interfaces vary with temperature.

% 5 K, 1970 nV
% 50 K, 540 nV
% 300 K, 160 nV
% The decay of spin currents with different NiO thickness confirms that spin current generated from the GGG and the bottom YIG layer can flow through the NiO spacer and the top YIG layer and finally reach the Pt strip.
% FIGURE 4
%\newpage

\section{Summary}
In summary, the spin transport properties of an insulating trilayer based on two ferrimagnetic insulators separated by a thin antiferromagnetic insulator were presented. The spin diffusion length of NiO was found to be $\lambda_{\mathrm{NiO}} \simeq 4.2\  \mathrm{nm}$ at room temperature. In addition, a large increase of the SSE signal was observed below 30~K, revealing the dramatic effects of paramagnetic SSE from the GGG substrate. The field dependence of the SSE shows the switching of YIG magnetization at low field as well as paramagnetic behavior associated with GGG. Furthermore, the SSE voltages show a broad peak around 100 K, a feature that may be related to the temperature dependence of spin diffusion length in YIG. This experimental study provides information on how spins can be generated, transported and detected in a heterostructure consisting of paramagnetic, ferrimagnetic and antiferromagnetic insulators.

\section*{Supplementary Material}
The supplementary material provides the details of sample characterization by scanning transmission electron microscopy (SEM), Vibrating Sample Magnetometer (VSM), transport measurements of a GGG/YIG/Pt sample, and field-dependent measurements of a GGG/YIG/NiO/YIG/Pt sample above 50 K.

\section*{Acknowledgements}
This work was supported partially by the MRSEC Program of the National Science Foundation under Award Number DMR-1420073. The instrumentation used in this research was support in part by the Gordon and Betty Moore Foundation¡¯s EPiQS Initiative through Grant GBMF4838 and in part by the  National Science Foundation under award NSF-DMR-1531664. ADK received support from the National Science Foundation under Grant No. DMR-1610416. JD and MW were supported by the U. S. National Science Foundation under Grants No. EFMA-1641989 and No. ECCS-1915849. Sandia National Laboratories is a multi-program laboratory managed and operated by National Technology and Engineering Solutions of Sandia, LLC., a wholly owned subsidiary of Honeywell International, Inc., for the U.S. Department of Energy's National Nuclear Security Administration under contract DE-NA-0003525. The views and conclusions contained herein are those of the authors and should not be interpreted as necessarily representing the official policies or endorsements, either expressed or implied, of the U.S. Government. We also would like to express our thanks to Pradeep Manandhar from Spin Memory Inc. for TEM characterization.

%\newpage
%\section{Bibliography}
%\clearpage

\end{document}